\documentclass[12pt]{article}

\begin{document}

\author{E. Ahmed$^{1}$, A. S. Elgazzar$^{2,3}$ and A. S. Hegazi$^{1}$ \\
$^{1.}$Mathematics Department, Faculty of Science\\
35516 Mansoura, Egypt\\
$^{2.}$Mathematics Department, Faculty of Education\\
45111 El-Arish, Egypt\\
$^{3.}$Mathematics Department, Faculty of Science - Al-Jouf\\
King Saud University, Kingdom of Saudi Arabia\\
E-mail: elgazzar@mans.edu.eg}
\title{An Overview of Complex Adaptive Systems}
\date{}
\maketitle

\begin{abstract}
Almost every biological, economic and social system is a complex
adaptive system (CAS). Mathematical and computer models are
relevant to CAS. Some approaches to modeling CAS are given.
Applications in vaccination and the immune system are studied.
Mathematical topics motivated by CAS are discussed.
\end{abstract}

\section{Basics of complex adaptive systems (CAS)}

\textbf{Definition (1):} A complex adaptive system consists of
inhomogeneous, interacting adaptive agents. Adaptive means capable
of learning.\newline
\newline
\textbf{Definition (2):} An emergent property of a CAS is a
property of the system as a whole which does not exist at the
individual elements (agents) level.\\
\\
Typical examples are the brain, the immune system, the economy, social
systems, ecology, insects swarm, etc..

Therefore to understand a complex system one has to study the system as a
whole and not to decompose it into its constituents. This totalistic
approach is against the standard reductionist one, which tries to decompose
any system to its constituents and hopes that by understanding the elements
one can understand the whole system.

\section{Why should we study complex adaptive systems?}

Most of living systems are CAS. Moreover they have intrinsic
unpredictability which causes some ''seemingly wise'' decisions to
have harmful side effects. Therefore we should try to understand
CAS to try to minimize such side effects. Here we give two
examples of these side effects.

Mathematical models have played important roles in understanding the impact
of vaccination programs. The complications of infectious diseases spread
make the problem of predicting the impact of vaccinations a nonlinear
problem. Sometimes a counter-intuitive result appears e.g. the threshold
phenomena [Edelstein-Keshet 1988]. Here another example will be mentioned.

Several vaccination programs are known e.g. mass vaccination where all
population is vaccinated, target vaccination where only a certain group is
vaccinated.

If one tries to understand the expected impact of a vaccination program one
should take the following points into account:

\begin{enumerate}
\item  Vaccination is not perfect hence a probability of vaccination failure
should be assumed.

\item  Sometimes vaccination takes time to be effective.

\item  Immunity is waning i.e. may be lost with time.

\item  Long range contacts can play a significant role e.g. SARS (severe
acute respiratory syndrome) has been transmitted between countries via air
travellers.
\end{enumerate}

Rubella is a mild viral infectious disease. Typically it is most dangerous
when infecting a pregnant female where it has severe effects on the fetus.
Once one gets it he (she) gets a life long immunity. There are several
vaccination strategies for rubella [Vynnycky et al 2003]. The US policy is
to vaccinate all two years old children. The UK policy is to vaccinate only
14-years old girls. Another strategy which is adopted in some underdeveloped
countries is not to vaccinate at all. It has been found [Jazbec et al 2003]
that in most cases the UK strategy is equal or better than the US one
despite being cheaper.

An interesting situation arose when some countries adopted a private sector
vaccination to MMR (Measles, Mumps and Rubella) [Vynnycky et al 2003]. It
was expected that the number of Congenital Rubella Syndrome (CRS) will
decrease. However it did not and in some countries (e.g. Greece and Costs
Rica) it increased. The reason can be understood as follows: This
vaccination to part of the population decreases the probability of
contracting the disease at young age. Hence the number of susceptible
individuals at adulthood increases. Consequently the probability of
contracting the disease at adulthood increases. This is an example of the
counterintuitive effects of some vaccination programs.

Another example for bad side effects is Lake Victoria [Chu et al 2003] where
a new species called Nile perch was introduced expecting that it is more
economically profitable. Yet the following results have appeared:

\begin{enumerate}
\item[(i)]  The local fishermen's tools were not suitable for the new fish
hence only large corporations benefited.

\item[(ii)]  Due to its higher price the locals were unavailable to buy the
new type.

\item[(iii)]  The original fish used to eat the larva of mosquitoes but now
mosquitoes' numbers have increased significantly thus the quality of life of
the locals have deteriorated!!
\end{enumerate}

There are at least two sources for unpredictability in CAS. The first is the
nonlinear interactions between its agents [West 1990]. The second is that
CAS are open systems hence perturbation to one system may affect another
related one e.g. perturbation to Lake Victoria affected the number of
mosquitoes.

\section{How to model a CAS?}

The standard approaches are

\begin{enumerate}
\item  Ordinary differential equations (ODE), difference equations and
partial differential equations (PDE).

\item  Cellular automata (CA) [Ilachinski 2001].

\item  Evolutionary game theory [Hofbauer and Sigmund 1998].

\item  Agent based models.

\item  Networks [Watts and Strogatz 1998] etc..

\item  Fractional calculus [Stanislavsky 2000].
\end{enumerate}

Some of these approaches are included in [Boccara 2004].

The ODE and PDE approaches have some difficulties as follows [Louzon et al
2003]:

\begin{enumerate}
\item[(i)]  ODE and PDE assumes that local fluctuations have been smoothed
out.

\item[(ii)]  Typically they neglect correlations between movements of
different species.

\item[(iii)]  They assume instantaneous results of interactions.
\end{enumerate}

Most biological systems show delay and do not satisfy the above assumptions.
They concluded that a cellular automata (CA) [Ilachinski 2001] type system
called microscopic simulation is more suitable to model complex biological
systems. We agree that CA type systems are more suitable to model complex
biological systems but such systems suffer from a main drawback namely the
difficulty of obtaining analytical results. The known analytical results
about CA type systems are very few compared to the known results about ODE
and PDE. Some mathematical results about CA are given in the appendix.

Now we present a compromise i.e. a PDE which avoids the delay and the
correlations drawbacks. It is called telegraph reaction diffusion equations
[Ahmed and Hassan 2000]. To overcome the non-delay weakness in Fick's law it
is replaced by

\begin{equation}
J(x,t)+\tau \frac{\partial J(x,t)}{\partial t}=-D\frac{\partial c}{\partial x%
},
\end{equation}
where the flux $J(x,t)$ relaxes, with some given characteristic
time constant $\tau $ and $c$ is the concentration of the
diffusing substance. Combining Eq. (1) with the equation of
continuity, one obtains the modified diffusion equation or the
Telegraph equation:

\begin{equation}
\frac{\partial c}{\partial t}+\tau \frac{\partial ^{2}c}{\partial x^{2}}=D%
\frac{\partial ^{2}c}{\partial x^{2}}.
\end{equation}
The corresponding Telegraph reaction diffusion (TRD) is given by

\begin{equation}
\tau \frac{\partial ^{2}c}{\partial t^{2}}+\left( 1-\frac{\mathrm{d}f(c)}{%
\mathrm{d}c}\right) \frac{\partial c}{\partial t}=D\frac{\partial ^{2}c}{%
\partial x^{2}}+f(c),
\end{equation}
where $f(c)$ is a polynomial in $c$.

Another motivation for TRD comes from media with memory where the
flux $J$ is related to the density $c(x,t)$ through a relaxation
function $K(t)$ as follows

\[
J(x,t)=-\int_{0}^{t}K(t-\acute{t})\frac{\partial c(x,\acute{t})}{\partial x}%
\mathrm{d}\acute{t}.
\]
It can be shown [Compte \& Metzler 1997] that, with a suitable choice for
the kernel $K(t)$, the standard Telegraph equation is obtained.

A third motivation is that starting from discrete space time one does not
obtain the standard diffusion equation but the telegraph equation [Chopard
and Droz 1991].

Moreover it is known that TRD results from correlated random walk [Diekmann
et al, 2000]. This supports the conclusion that Telegraph reaction diffusion
equation is more suitable for modeling complex systems than the usual
diffusion one.

\section{The immune system as a complex system [Segel and Cohen 2001, Ahmed
and Hashish 2004]}

The emergent properties of the immune system (IS) included:

\begin{enumerate}
\item[{*}]  The ability to distinguish any substance (typically
called antigen Ag) and determine whether it is damaging or not. If
Ag is non-damaging (damaging) then, typically, IS tolerates it
(responds to it).

\item[{*}]  If it decides to respond to it then IS determines whether to
eradicate it or to contain it.

\item[{*}]  The ability to memorize most previously encountered Ag, which
enables it to mount a more effective reaction in any future encounters. This
is the basis of vaccination processes.

\item[{*}]  IS is complex thus it has a network structure.

\item[{*}]  The immune network is not homogeneous since there are effectors
with many connections and others with low number of connections.

\item[{*}]  The Ag, which enters our bodies, has extremely wide diversity.
Thus mechanisms have to exist to produce immune effectors with constantly
changing random specificity to be able to recognize these Ag. Consequently
IS is an adaptive complex system.

\item[{*}]  Having said that, one should notice that the wide diversity of
IS contains the danger of autoimmunity (attacking the body). Thus mechanisms
that limit autoimmunity should exist.

\item[{*}]  In addition to the primary clonal deletion mechanism,
two further brilliant mechanisms exist: The first is that the IS
network is a threshold or "window" one i.e. no activation exists
if the Ag quantity is too low or too high (This is called low and
high zone tolerance).

\item[{*}]  Thus an auto reactive immune effector (i.e. an immune effector
that attacks the body to which it belongs) will face so many self-antigens
that it has to be suppressed due to the high zone tolerance mechanism.

\item[{*}]  Another mechanism against autoimmunity is the second
signal given by antigen presenting cells (APC). If the immune
effector is self reactive then, in most cases, it does not receive
the second signal thus it becomes anergic.

\item[{*}]  Also long term memory can be explained by the phenomena of high
and low zone tolerance where IS tolerates Ag if its quantity is too high or
too low. So persisting Ag is possible and continuous activation of immune
effectors may occur.

\item[{*}]  There is another possible explanation for long term
memory using the immune system (Extremal Dynamics).

\item[{*}]  Thus design principles of IS can explain important phenomena of
IS.
\end{enumerate}

An interesting example is given by Matzinger [Matzinger 2002] where she
argued that to prevent transplant rejection it may be more useful to design
drugs that blocks signal II and not signal I (which the present drugs do).
The reason is blocking signal II make the effectors (which originally were
capable of recognizing the transplant) anergic while leaving the other
immune effectors intact.

\section{Conclusions}

\begin{enumerate}
\item[(i)]  CAS should be studied as a whole hence reductionist point of
view may not be reliable in some cases.

\item[(ii)]  CAS are open with nonlinear local interactions hence:

\begin{enumerate}
\item[1.]  Long range prediction is highly unlikely [Strogatz 2000, Holmgren
1996].

\item[2.]  When studying a CAS take into consideration the effects of its
perturbation on related systems e.g. perturbation of lake Victoria has
affected mosquitoes' numbers hence the locals quality of life. This is also
relevant to the case of natural disasters where an earthquake at a city can
cause a widespread power failure at other cities.

\item[3.]  Expect side effects to any ''WISE'' decision.

\item[4.]  Mathematical and computer models may be helpful in reducing such
side effects.
\end{enumerate}

\item[(iii)]  Optimization in CAS should be multi-objective and not single
objective [Collette and Siarry 2003].

\item[(iv)]  CAS are very difficult to control. Interference at
highly connected sites may be a useful approach [Dorogovtsev and
Mendez 2004]. The interlinked nature of CAS elements complicates
both the unpredictability and controllability problems. It also
plays an important role in innovations spread.

\item[(v)]  Memory effects should not be neglected in CAS. This lends more
support for the proposed telegraph reaction diffusion Eq. (3). Also memory
games have been studied [Smale 1980, Ahmed and Hegazi 2000]. Also delay and
fractional calculus are relevant to CAS.

\item[(vi)]  Mathematical topics motivated by CAS include ODE and PDE
(non-autonomous, delayed, periodic coefficients, stability and persistence),
multi-objective optimization (including biologically motivated methods e.g.
Ant colony optimization, Extremal optimization, Genetic algorithm etc),
difference equations, cellular automata, networks, fractional calculus,
control (e.g. bounded delayed control of distributed systems), game theory,
nonlinear dynamics and fuzzy mathematics.
\end{enumerate}

Some of the mathematics motivated by CAS will be reviewed in the
appendices.

\section*{Acknowledgments}

One of the authors (A. S. Hegazi) acknowledge the financial
support of a Mansoura University grant.

\section*{References}

\begin{enumerate}
\item[ ]  Ahmed E. and Hassan S.Z. (2000), "On Diffusion in Some
Biological and Economic Systems", Z. Naturforsch. 55a, 669.

\item[ ]  Ahmed E. and Hashish A.H. (2004), "On Modeling the
immune system as a complex system", Theor. BioSci. (Accepted).

\item[ ]  Ahmed E. and Hegazi A.S."On Discrete Dynamical Systems
Associated to Games", Adv. Complex. Sys. 2, 423-429 (2000).

\item[ ]  Ahmed E., Hegazi A.S. "On Persistence and Stability of
Some Spatially Inhomogeneous Systems", J. Math. Analys. Appl. 268,
74-88 (2002).

\item[ ]  Ahmed E., Hegazi A.S., Elgazzar A.S. and Yehia H.M., "On
Synchronization, Persistence and Seasonality in Some Inhomogeneous
Models in Epidemics and Ecology", Physica A 322, 155-168 (2003).

\item[ ]  Ahmed E. Hegazi A.S. and Abdel-Hafiz A.T., "On
Multiobjective oligopoly" Nonlin. Dyn. Psych. Life Sci. 7, 203-217
(2003).

\item[ ]  Ahmed, E. and Elgazzar, A.S., "On Coordination and Continuous 
Hawk-Dove Games on Small-World Networks", Eur. Phys. J. B 18, 159 (2000 a).

\item[ ]  Ahmed, E. and Elgazzar, A.S., "On the Dynamics of Local Hawk-Dove 
Game", Int. J. Mod. Phys. C 11, 607 (2000 b).

\item[ ]  Ahmed, E., Hegazi, A.S. and Elgazzar, A.S., "An Epidemic Model on 
Small-World Networks and Ring Vaccination", Int. J. Mod. Phys. C 13, 189 (2002).

\item[ ]  Albert, R. and Barab\'asi, A.L., "Statistical mechanics of complex 
networks", Rev. Mod. Phys. 74, 47 (2002).

\item[ ]  Bagnoli F., Boccara N. and Rechtman R.(2002), "Nature of
phase transitions in probabilistic cellular automata with two
absorbing states", Lect. notes Computer sci. 2493, 249-258.

\item[ ]  Barnett S. (1990),"Matrices",Oxford Univ. Press, U.K.

\item[ ]  Boccara N. (2004), "Modeling complex systems", Springer
Publ., Berlin.

\item[ ]  Chopard B. and Droz M. (1991), "Cellular automata model
for the diffusion equation", J. Stat. Phys. 64, 859.

\item[ ]  Chu D., Strand R. and Fjelland R. (2003), "Theories of
Complexity", Complexity 8, 19.

\item[ ]  Collette Y. and Siarry P. (2003), "Multiobjective
Optimization", Springer, Berlin.

\item[ ]  Compte A. and Metzle R.(1997), "The generalized Cattaneo
equation for the description of anomalous transport processes" J.
Phys. A 30, 7277.

\item[ ]  Diekmann O., R. Durrett, K. P. Hadeler, P. Maini, H. L.
Smith, V. Capasso (eds.) (2000), "Mathematics inspired by
biology", Springer Verlag, Berlin.

\item[ ]  Dorogovtsev S.N. and Mendes J.F (2004), "The shortest
path to complex networks", Cond-mat 0404593.

\item[ ]  Domany E. and Kinzel W.(1984), Equivalence of CA and Ising models
and directed percolation, Phys.Rev.Lett.53, 311.

\item[ ]  Edelstein-Keshet L. (1988), "Introduction to
mathematical biology", Random House, N.Y.

\item[ ]  Elgazzar, A.S., "Application of the Sznajd Sociophysics Model 
to Small-World Networks", Int. J. Mod. Phys. C 12, 1537 (2001).

\item[ ]  Elgazzar, A.S., "A Model for the Evolution of Economic Systems 
in Social Networks", Physica A 303, 543 (2002).

\item[ ]  Erd\"os, P. and R\'enyi, A., "On the Evolution of Random Graphs", Publ. Math. Inst. Hung.
Acad. Sci. 5, 17 (1960).

\item[ ]  Holmgren R. (1996), "A first course in discrete
dynamical systems", Springer, Berlin.

\item[ ]  Hofbauer J. and Sigmund K. (1998), "Evolutionary games
and population dynamics", Cambridge Univ. Press, U.K.

\item[ ]  Ilachinski A.(2001), "Cellular automata", World
Scientific Publ., Singapore.

\item[ ]  Jazbec A., Delimar M. and Vrzic V.S. (2003), "Simulation
model of Rubella", Appl. Math. Comp. (To Appear).

\item[ ]  Jen E.(1990), "Aperiodicity in 1-dimensional cellular
automata", Physica D 45, 3.

\item[ ]  Kaneko K. (1993), "Theory and application of coupled map
lattices," Wiley Pub.

\item[ ]  Louzoun Y., Solomon S., Atlan H. and Cohen I.R. (2003),
"Proliferation and competition in discrete biological systems",
Bull. Math. Biol. 65, 375.

\item[ ]  Martin O., Odlyzko A.M. and Wolfram S. (1984),
"Algebraic properties of cellular automata", Comm. Math. Phys. 93,
219.

\item[ ]  Matzinger P. (2002), "The danger model", Science 296,
12, pp. 301.

\item[ ]  Milgram, S., "The Small World Problem", Psychol. Today 2, 60 (1967).

\item[ ]  Rocco A. and West B.J. (1999), "Fractional Calculus and
the Evolution of Fractal Phenomena", Physica A 265, 535.

\item[ ]  Schonfisch B. and De Roos A.(1999), "Synchronous and
asynchronous updating in cellular automata", Biosys. 51, 123-143.

\item[ ]  Segel L.A. and Cohen I.R.(eds.)(2001), "Design
principles for the immune system and other distributed autonomous
systems", Oxford Univ. Press, U.K.

\item[ ]  Smale S.(1980). "The prisoners' dilemma and dynamical
systems associated to games", Econometrica 48, 1617-1634.

\item[ ]  Smith J.B.(2003), "Complex systems", CS 0303020.

\item[ ]  Stanislavsky A.A. (2000), "Memory effects and
macroscopic manifestation of randomness", Phys. Rev. E 61, 4752.

\item[ ]  Stevens J.G., Rosenzweig R.E. and Cerkanowicz
A.E.(1993), "Transient and cyclic behavior of cellular automata
with null boundary conditions", J. Stat. Phys. 73, 159-174.

\item[ ]  Strogatz S. (2001), "Nonlinear dynamics and chaos",
Perseus Books Group

\item[ ]  Tadaki S. (1994), "Periodicity of cylindrical cellular
automata, Nonlin. 9412003

\item[ ]  Vynnycky E., Gay N.J. and Cutts F.T. (2003), "The
predicted impact of private sector MMR vaccination on the burden
of Congenital Rubella Syndrome", Vaccine 21, 2708.

\item[ ]  Watts D.J. and Strogatz S.H. (1998), "Collective
dynamics of small world network", Nature 393, 440.

\item[ ]  West B.J. (1991), "Fractal Physiology and chaos in
medicine", World Scientific Publ., Singapore.
\end{enumerate}

\section*{Appendix (1): Some mathematical results for one-dimensional cellular automata}

\textbf{Definition (3):} A cellular automata consists of 4 components: A
graph $G$, a set of states such that each site (vertex) of the graph has one
of the possible states, a neighborhood set which assigns to each vertex a
certain neighborhood and a transition function $f$ which defines the
evolution of the state of each site as a function of the states of that site
and those in its neighborhood.\newline
\newline
We choose the set of possible states to be the ring $Z(p)$ i.e. the set of
integers $0,1,2,...,p-1$, where addition is defined $\mathrm{mod\;}p$. The
total number of sites is denoted by $N$. In most of the cases, we choose $%
N,\;p$ to be relatively prime. The set of states of the sites at a
given time is called a configuration. We now restrict us to a
one-dimensional space. Let $x(j,t)$ be the state of site $j$ at
time $t$.\newline
\newline
\textbf{Definition (4):} A finite initial configuration is one such that
there are two natural numbers $L,\;R$ such that $0<L<R<N$, and $x(j,0)=0$ if
$j<L$ or $j>R$.\newline
\newline
\textbf{Theorem (1)} [Jen 1990]: If $x(i,t),\;x(j,t),\;i<j$ are two periodic
sequences i.e. $x(i,t)=x(i,t+p(i)),\;x(j,t)=x(j,t+p(j))$, then for every $k$
such that $i<k<j$ then $x(k,t)$ is periodic.\newline
\newline
\textbf{Corollary (1)} [Jen 1990]: If CA evolves according to the rule
\begin{equation}
x(i,t)=f(x(i-1,t),x(i,t),x(i+1,t))\;\mathrm{mod\;}2,
\end{equation}
such that $000\rightarrow 0,100\rightarrow 0,001\rightarrow 0$, then for any
finite initial configuration the system is temporarily periodic i.e. the
sequence $(x(i,t))$ is periodic for all $i$ such that $0<i\leq N,0<T<t$.%
\newline
\newline
\textbf{Proof.} The fact that $100\rightarrow 0$ implies that $x(i,t)=0$ for
$i>R$, similarly $x(i,t)=0$ for $i<L$ for all $t>0$. Applying theorem (1)
the result is proved.\newline

In the case that $f$ in Eq. (4) is linear, one can use the methods of
[Stevens et al 1993, Tadaki 1994] to get useful information about possible
periodicity's of the system. In this case the system can be written as
\begin{equation}
X(t+1)=UX(t),
\end{equation}
where $U$ is called the evolution matrix. Then $X(t)=U^{t}X(0)$. In this
case the asymptotic behavior of the system is governed by the characteristic
polynomial of $U$ on the field $Z(p)$. Assuming periodic boundary
conditions, the matrix $U$ is circulant matrix [Barnett 1990].

Let $P(N,\lambda )$ be the characteristic polynomial of the system (5) with $%
N$ sites, then typically it has the form
\begin{equation}
P(n,\lambda )=\lambda ^{a}d(n,\lambda ),\;d(n,0)=1.
\end{equation}
If $a>0$, then the systems tends to a fixed configuration (which corresponds
to a fixed point for discrete time continuous state dynamical systems).
Reducing $d(n,\lambda )$ to its irreducible factors on the field of states
then in most cases a cycle of length $p^{k}-1$ exist for the system where $k$
is the degree of the irreducible factors.

As an example consider rule 90 [Martin et al 1984]
\begin{equation}
x(i,t+1)=x(i-1,t)+x(i+1,t)\;\mathrm{mod\;}2.
\end{equation}
For $N=5$, we have $P(5,\lambda )=\lambda (\lambda ^{2}+\lambda +1)^{2}\;%
\mathrm{mod\;}2$, hence the system may evolve to a fixed configuration (e,g,
$x(i,t)=0$ for all $t>T>0$, for all $0<i\leq N$). It can also evolve to a
cycle of period $3\;(=2^{2}-1)$.

Similarly for $N=9$, $P(9,\lambda )=\lambda (\lambda +1)^{2}(\lambda
^{3}+l+1)^{2}$ on $Z(2)$. Hence this system may evolve to a fixed
configuration or to a periodic one with period 7. For $N=13$, similar study
implies that $P(13,\lambda )=\lambda (\lambda ^{6}+\lambda ^{5}+\lambda
^{4}+\lambda +1)^{2}$ on $Z(2)$. Hence fixed configurations and periodic
ones with period 63 are expected. Such long periods may not be easy to find
numerically. These results can be obtained using more elaborate methods
[Martin et al 1984]; but the simplicity of the present approach is appealing.

Moreover it is directly applicable to nonlocal cases which have gained much
attention after the pioneering work of Watts and Strogatz on small world
network (SWN) [Watts and Strogatz 1998]. As an example consider the
following system
\begin{equation}
x(i,t+1)=x(i-1,t)+x(i+1,t)+x(i+k,t)\;\mathrm{mod\;}2,
\end{equation}
where $k$ is fixed. Some of the characteristic polynomials P(N,$\lambda $,k)
are:
\begin{eqnarray}
P(11,0,\lambda ) &=&\lambda ^{11}+\lambda ^{10}+\lambda
^{5}+\lambda
^{4}+\lambda +1,  \nonumber \\
P(11,3,\lambda ) &=&\lambda ^{11}+\lambda ^{9}+\lambda
^{7}+\lambda
^{6}+\lambda ^{5}+\lambda ^{4}+\lambda +1,  \nonumber \\
P(11,1,\lambda ) &=&\lambda ^{11}+\lambda ^{8}+\lambda
^{7}+\lambda ^{5}+\lambda ^{2}+1,
\end{eqnarray}
Hence we have the following proposition:\newline
\newline
\textbf{Proposition (1):} a) The system (8) depends on $k$.\newline
\newline
b) The asymptotic behavior of (8) contains the following: For $N=11,\;k=3$,
no fixed configuration but a periodic one with period 1023.\newline
\newline
\textbf{Proof.} a) For $N=11,\;k=5$, a homogeneous configuration
is expected. This is not the case for $N=11,\;k=0$ or
$k=3$.\newline
\newline
b) Use the procedure explained before.\newline

Typically updating of CA is synchronous. It is important to notice that
other types of updating e.g. a uniform random asynchronous one (where only
one site is chosen randomly and updated at each time step) gives other
patterns [Schonfisch and de Roos 1999]. The following lemma is useful\newline
\newline
\textbf{Lemma (1)}: a) States which are stationary under synchronous
updating are also stationary under asynchronous one.\newline
\newline
b) If there is a site $j$ which is not updated for all time $t>T>0$ then
stationary configuration with respect to asynchronous updating may not be so
under synchronous one.\newline
\newline
\textbf{Proof.} a) If $f(x(1),x(2),...,x(N))=(x(1),x(2),...,x(N))$ then $%
f(j,x(j))=x(j)$. This proves part a). Since site $j$ is not updated for $%
t>T>0$ then $f(j,x(j))\neq x(j)$ can still belong to a homogeneous
configuration for the asynchronous updating but not the homogeneous one.
This proves b).\newline

Loosely speaking patterns present in asynchronous updating are mostly
present in synchronous one. Motivated by these results we study sequential
CA e.g. the sequential rule 90 is
\begin{equation}
x(j,t+1)=x(j-1,t+1)+x(j+1,t)\;\mathrm{mod\;}2.
\end{equation}
This can be written in the following equivalent form
\begin{equation}
x(j,t+1)=\sum_{k=2}^{j+2}x(k,t)\;\mathrm{mod\;}2,
\end{equation}
where free periodic boundary conditions are assumed. The characteristic
polynomials of the system (7) are:
\begin{eqnarray}
P(5,\lambda ) &=&\lambda ^{5}+\lambda ^{3},\;P(6,\lambda )=\lambda
^{6}+\lambda ^{5}+\lambda ^{3},\;P(7,\lambda )=\lambda ^{7},  \nonumber \\
P(13,\lambda ) &=&\lambda ^{7}(\lambda ^{3}+\lambda ^{2}+1)^{2}.
\nonumber
\end{eqnarray}
Hence homogeneous configurations are expected for $N=5,6,7$. For $N=13$ a
periodic configuration with period 7 is expected.

Studying the system (10) numerically showed that chaos (in the sense of
sensitive dependence on initial conditions which is sometimes called damage
spread) exists.\newline
\newline
\textbf{Proposition (2)}: Every initially finite configuration will evolve
under the CA
\begin{equation}
x(j,t+1)=x(j-1,t+1)\;x(j+1,t)\;\mathrm{mod\;}2,
\end{equation}
into the zero configuration $x(j,t)=0$ for all $j$, $0<j\leq N$, for all
time $t>T>0$ where $T<N$.\newline
\newline
\textbf{Proof.} We have
\[
x(R,1)=x(R+1,0)\;x(R-1,1).
\]
But $x(R+1,0)=0$ by definition of initially finite configuration thus $%
x(R,1)=0$. Repeating for $x(R-1,2)$, one gets $x(R-1,2)=0$ and continue.%
\newline

Now the above results are applied to two known examples. The first is
Domany-Kinzel (DK) model [Kinzel and Domany 1984], which is given by:
\begin{eqnarray}
\mathrm{If\;}x(j-1,t)+x(j+1,t) &=&0\;\mathrm{then\;}x(j,t+1)=0.  \nonumber \\
\mathrm{If\;}x(j-1,t)+x(j+1,t) &=&1\;\mathrm{then\;}x(j,t+1)=1,  \nonumber \\
\mathrm{with}\;\mathrm{probability\;}p_{1}. \\
\mathrm{If\;}x(j-1,t)+x(j+1,t) &=&2\;\mathrm{then\;}x(j,t+1)=1,  \nonumber \\
\mathrm{with\;probability\;}p_{2}.  \nonumber
\end{eqnarray}
where $x(j,t)$ are Boolean variables. For $p_{1}\rightarrow 1$, $%
p_{2}\rightarrow 1$, the system (13) corresponds to the CA
\begin{equation}
x(j,t+1)=x(j-1,t)+x(j+1,t)+x(j-1,t)\;x(j+1,t)\;\mathrm{mod\;}2.
\end{equation}
\newline
\newline
\textbf{Proposition (3):} Any finite initial configuration with two
consecutive ones will tend to the homogeneous configuration $x(j,t)=1$ for
all $0\leq j\leq N$, $0<T<t$, $T$ is sufficiently large under the CA (12).
Consequently the region $p_{1}\rightarrow 1$, $p_{2}\rightarrow 1$ in the DK
CA does not show chaos (damage spread).\newline
\newline
\textbf{Proof.} Assume that $x(j,0)=x(j+1,0)=1$. Then the system (14)
implies
\[
x(j-1,1)=x(j,1)=x(j+1,1)=x(j+2,1)=1.
\]
Continue one gets after $t$ time steps $x(k,t)=1$, where $j-t\leq k\leq
j+t+1 $. This proves the first part. Now since the CA (14) will tend to $%
x(j,t)=1$ for all $0\leq j\leq N$, $0<T<t$, then any change in the initial
conditions that preserves the condition $x(j,0)=x(j+1,0)=1$ for some $j$
will not affect the asymptotic behavior of the CA (14). This completes the
proof.\newline

The case $p_{1}\rightarrow 0$ in the DK model corresponds to the CA
\begin{equation}
x(j,t+1)=x(j-1,t)\;x(j+1,t)\;\mathrm{mod\;}2.
\end{equation}
Following similar steps as those in proposition (3) one can prove the
following:\newline
\newline
\textbf{Proposition (4):} Any finite initial configuration with two
consecutive zeros will tend to the homogeneous configuration $x(j,t)=0$ for
all $0\leq j\leq N$, $0<T<t$, $T$ is sufficiently large under the CA (15).
Consequently the region $p_{1}\rightarrow 1,\;p_{2}\rightarrow 1$ in the DK
CA does not show chaos (damage spread) or periodic configurations.\newline
\newline
In the limit $p_{2}\rightarrow 0,\;p_{1}\rightarrow 1$, DK model corresponds
to rule 90
\[
x(j,t+1)=x(j-1,t)+x(j+1,t)\;\mathrm{mod\;}2,
\]
which is known to be chaotic.

All of the above results agree with numerical simulations. Bagnoli
et al model [Bagnoli et al 2002] is given by
\begin{eqnarray}
\mathrm{If}\;x(j-1,t)+x(j,t)+x(j+1,t) &=&0\;\mathrm{then\;}x(j,t+1)=0.
\nonumber \\
\mathrm{If\;}x(j-1,t)+x(j,t)+x(j+1,t) &=&1,  \nonumber \\
\mathrm{then\;}x(j,t+1) &=&1\;\mathrm{with\;probability\;}p_{1}.  \nonumber
\\
\mathrm{If\;}x(j-1,t)+x(j,t)+x(j+1,t) &=&2, \\
\mathrm{then\;}x(j,t+1) &=&1\;\mathrm{with\;probability\;}p_{2}.  \nonumber
\\
\mathrm{If\;}x(j-1,t)+x(j,t)+x(j+1,t) &=&3\;\mathrm{then\;}x(j,t+1)=1.
\nonumber
\end{eqnarray}
where $x(j,t)$ are Boolean variables. The limit $p_{1}\rightarrow
1,\;p_{2}\rightarrow 1$ corresponds to the CA
\begin{eqnarray}
x(j,t+1) &=&x(j-1,t)+x(j+1,t)+x(j,t)+x(j-1,t)\;x(j+1,t)+  \nonumber \\
&&x(j,t)\;x(j+1,t)+x(j-1,t)\;x(j,t)+ \\
&&x(j-1,t)\;x(j,t)\;x(j+1,t)\;\mathrm{mod\;}2.  \nonumber
\end{eqnarray}
The limit $p_{1}\rightarrow 0,\;p_{2}\rightarrow 0$ corresponds to the CA
\begin{equation}
x(j,t+1)=x(j-1,t)\;x(j,t)\;x(j+1,t)\;\mathrm{mod\;}2.
\end{equation}
\newline
\textbf{Proposition (5):} a) Any nonzero finite initial configuration will
evolve under the CA (17) into the homogeneous configuration $x(j,t)=1$ for
all $0\leq j\leq N$, for $t$ is sufficiently large. Hence the limit $%
p_{1}\rightarrow 1,\;p_{2}\rightarrow 1$ in Bagnoli et al model does not
show chaos or periodic configurations.\newline
\newline
b) Any finite initial configuration containing at least one zero site will
evolve under the CA (18) into the homogeneous configuration $x(j,t)=0$ for
all $0\leq j\leq N$, for $t$ is sufficiently large. Hence the limit $%
p_{1}\rightarrow 0,\;p_{2}\rightarrow 0$ in Bagnoli et al model does not
show chaos or periodic configurations.\newline
\newline
\textbf{Proof.} similar to proposition (3).\newline

The limit $p_{1}\rightarrow 1,\;p_{2}\rightarrow 0$ corresponds to the CA
\begin{eqnarray}
x(j,t+1) &=&x(j-1,t)+x(j+1,t)+x(j,t)+  \nonumber \\
&&2x(j-1,t)\;x(j,t)\;x(j+1,t)\;\mathrm{mod\;}2
\end{eqnarray}
which is similar to rule 150 $x(j,t+1)=x(j-1,t)+x(j+1,t)+x(j,t)$, hence
chaos is expected in Bagnoli et al model in this limit. All of the above
results agree with numerical simulations.

It is interesting how CA unite polynomials on finite fields, circulant
matrices, graph theory techniques and many other branches of mathematics
into one branch which is important both mathematically and from the point of
view of applications in complex systems.

\section*{Appendix (2): Overview of networks in CAS}

Complex systems are often modeled as graphs where agents are the
vertices and the interactions form the edges of the graph.
Typically graphs are either regular lattices (e.g. square or
cubic), random or scale free where the probability that a vertex
has degree $k$ is $p(k)\approx k^{-\gamma }$. Most of the real
networks are of the scale free type. Some proposed mechanisms for
this fat tailed distribution [Dorogovtsev and Mendes 2004] are
self organization (c.f. biological systems) and optimization
involving many agents (c.f. economy).

Random graphs were first studied by the mathematicians Erd\"os and
R\'enyi [Erd\"os and R\'enyi 1960]. Their model consists of $N$
nodes, such that every pair of nodes is connected by a bond with
probability $p$. The recent increase in computing power and the
appearance of interdisciplinary sciences has lead to a better
understanding of the properties of complex networks.

Two main properties of complex networks are clustering and small world
effect.

Small-world effect means the average shortest node to node (vertex
to vertex) distance is very short compared with the whole size of
the system (total number of vertices). For social networks, the
social psychologist Milgram [Milgram 1967] concluded that the
average length of the path of acquaintances connecting each pair
of people in the United States is six. This concept is known as
the six degrees of separation. Such an effect makes it easier for
an effect (e.g. an epidemic) to spread throughout the network.

In a regular 1-dimensional lattice of size $N$, the average shortest path
connecting any two vertices $l$ increases linearly with the system size. So
regular lattices do not display small-world effect. On the other hand for a
random graph, with coordination number $z$, one has $z$ first (nearest)
neighbors, $z^{2}$ second neighbors and so on. This means that the total
number of vertices $N=z^{l}$, this gives
\[
l=\frac{\mathrm{Ln}(N)}{\mathrm{Ln}(z)}.
\]
The logarithmic increase with the size of the lattice allows the distance $l$
to be very short even for large $N$. Then random graphs display the
small-world effect.

Clustering is a common property of complex networks. It means that every
vertex has a group of connected nearest neighbours (NN) (collaborators,
friends), some of them will often be a connected NN to another vertex. As a
measure for the clustering property, a clustering coefficient $C$ is defined
as the probability that connected pairs of NN of a vertex are also connected
to each others. For a random graph, $C=z/N$ which goes to zero for large $N$%
. So random graphs do not display clustering property. On the other hand, a
fully connected regular lattice itself forms a cluster, then its cluster
coefficient is equal to $1$.

Complex networks display a small-world effect like random graphs,
and they have large clustering coefficient as regular lattices.
For a review on many real-world examples, see [Dorogovtsev and
Mendes 2004].

A small-world network (SWN) proposed initially by Watts and Strogatz [Watts
and Strogatz 1998] is a superposition of a regular lattice (with high
clustering coefficient) and a random graph (with the small world effect).
SWN satisfy the main properties of social networks. Also, the structure of
SWN combines between both local and nonlocal interactions which is observed
in many real systems. For example epidemic spreading show nonlocal
interactions e.g SARS.

The concept of SWN has been applied successfully in modelling many
CAS, e.g, some games [Ahmed and Elgazzar 2000 a], epidemics [Ahmed
et. al. 2002], economic systems [Elgazzar 2002], and opinion
dynamics [Elgazzar 2001].

An important property related to disease spread in a network is the second
moment of the degree distribution i.e. $\langle k^{2}\rangle $. If it is
divergent then on average a vertex has an infinite number of second nearest
neighbors thus if a single vertex is infected the disease will spread in the
whole network. This explains the results that disease spread on scale free
networks has zero threshold (contrary to the ODE and PDE models). However
one should realize that real networks are finite hence a kind of threshold
is expected.

Scale-free networks [Albert and Barab\'asi 2002] are another class
of complex networks. A scale-free network does not have a certain
scale. Some nodes have a huge number of connections to other
nodes, whereas most nodes have only a few, following a power law
distribution.

\section*{Appendix (3): Basics of game theory}

Game theory [Hofbauer and Sigmund 1998] is the study of the ways in which
strategic interactions among rational players produce outcomes (profits)
with respect to the preferences of the players. Each player in a game faces
a choice among two or more possible strategies. A strategy is a
predetermined program of play that tells the player what actions to take in
response to every possible strategy other players may use. A basic property
of game theory is that one's payoff depends on the others' decisions as well
as his.

The mathematical framework of the game theory was initiated by von Neumann
and Morgenstern in 1944. Also they had suggested the max-min solution for
games which is calculated as follows: Consider two players A and B are
playing against each other. Two strategies $S_{1}$, $S_{2}$ are allowed for
both of them. This game is called two-player, two-strategy game. Assume that
the constants $a,b,c$ and $d$ represent the payoffs (profits) such that, if
the two players use the same strategy $S_{1}(S_{2})$, their payoff is $a(d)$%
. When a player with strategy $S_{1}$ plays against another one with
strategy $S_{2}$, the payoff of the $S_{1}$-player is $b$ and the payoff of
the $S_{2}$-player is $c$ and so on. This is summarized in the payoff matrix
as follows:

\[
\begin{array}{ccc}
& S_{1} & S_{2} \\
S_{1} & a & b \\
S_{2} & c & d
\end{array}
.
\]
The max-min solution of von Neumann and Morgenstern is for the first player
to choose max\{min$(a,b)$,{min}$(c,d)$\}. The second player chooses {min}\{{%
max}$(a,c)$, {max}$(b,d)$\}. If both quantities are equal then the game is
stable. Otherwise use mixed strategies.

A weakness of this formalism has been pointed out by Maynard Smith in the
hawk-dove (HD) game whose payoff matrix is
\[
\Pi =
\begin{array}{ccc}
& \mathrm{H} & \mathrm{D} \\
\mathrm{H} & \frac{1}{2}(v-c) & v \\
\mathrm{D} & 0 & \frac{v}{2}
\end{array}
.
\]
The max-min solution implies (for $v<c$) that the solution is D yet as he
pointed out this solution is unstable since if one of the players adopts H
in a population of D he will have a very large payoff which will make other
players switch to H and so on till number of H is large enough that they
play each other frequently and get the low payoff $(v-c)/2$. Thus the stable
solution is that the fraction of hawks should be nonzero. To quantify this
concept one may use the replicator equation which intuitively means that the
rate of change of the fraction of players adopting strategy $i$ is
proportional to the difference between their payoff and the average payoff
of the population i.e.
\begin{equation}
\frac{\mathrm{d}x_{i}}{\mathrm{d}t}=x_{i}\left[ (\Pi x)_{i}-x\Pi x\right]
,\;i=1,2,...,n,\;\sum_{i=1}^{n}x_{i}=1,
\end{equation}
where $x_{i}$ is the fraction of players adopting strategy $i$, and $\Pi $
is the payoff matrix. Applying Eq. (20) to the HD game, one gets that the
asymptotically stable equilibrium solution is $x=v/c$, where $x$ is the
fraction of hawks in the population.

For asymmetric game the replicator dynamics equation is
\[
\frac{\mathrm{d}x_{i}}{\mathrm{d}t}=x_{i}\left[ (\Pi _{1}y)_{i}-x\Pi
_{1}y\right] ,\;\frac{\mathrm{d}y_{i}}{\mathrm{d}t}=y_{i}\left[ (\Pi
_{2}x)_{i}-y\Pi _{2}x\right] ,\;i=1,2,...,n.
\]
A basic drawback of normal game theory is the assumption that all
players interact globally. It is more realistic to study local
games [Ahmed and Elgazzar 2000 b] e.g. games on a lattice where
players interact only with their nearest neighbors. Also there are
several modifications for game formulations.

\section*{Appendix (4): Unpredictability in CAS}

There are at least two sources for unpredictability in CAS. The first is
that CAS are open systems hence perturbing a CAS may affect another related
one e.g. the insect population affected by the perturbation of Lake
Victoria. Another reason is the nonlinear interactions [Strogatz 2000]
between the elements of the CAS. The scientific and mathematical study of
Chaos Theory contains many overlaps with the study of Complex Systems, but
with differences related to method: Chaos Theory can be used to study
Complex Systems, but is not restricted to the study of these systems. Chaos
Theory ''deals with deterministic systems whose trajectories diverge
exponentially over time'' (Bar Yam, NECSI website). It has been used to
study Complex Systems, because these systems can be generally defined as a
''deterministic system that is difficult to predict''. On the other hand,
complexity deals with systems composed of many interacting agents'' The
point being that Chaos Theory is one of many tools and methods that can be
applied to the study of Complex Systems, but is not specifically devoted to
the way these systems are designed, developed, studied, and modeled. That
being stated, the famous example of the ''Butterfly Effect'' in a chaotic
system is an example of an agent (a butterfly) evoking a non-linear response
(the storm in New England) within a Complex System (Global Weather System).

A simple example of nonlinear interactions is the logistic difference
equation
\begin{equation}
x_{t+1}=rx_{t}(1-x_{t}),\;t=0,1,2,...,n,\;r>0.
\end{equation}
This equation has two equilibrium solutions $x=0,\;x=1-1/r\;(r>1)$ which are
asymptotically stable if $r<1$ or $1<r<3$ respectively. If $3<r<3.6$ then
cycles appears and if $r>3.6$ chaos sets in. Intuitively chaos is sensitive
dependence on initial conditions (for more mathematical definition see
[Holmgren 1996]). Hence in chaotic systems one cannot make long range
predictions c.f. weather. A useful measure of chaos are Lyapunov exponents
\begin{equation}
\lambda =\frac{1}{n}\sum_{t=0}^{n-1}\mathrm{Ln}\left| \acute{f}%
(x_{t})\right| .
\end{equation}
Since CAS consists of several interacting agents one studies coupled systems
e.g. coupled map lattices [Kaneko 1993] given by
\begin{equation}
x_{i}^{t+1}=(1-D)f(x_{i}^{t})+\frac{D}{2}\left[
f(x_{i-1}^{t})+f(x_{i+1}^{t})\right] ,\;i=1,2,...,n.
\end{equation}
The homogeneous equilibrium is given by $x=f(x)$ and it is asymptotically
stable if [Ahmed and Hegazi 2002]
\begin{equation}
\left| \acute{f}(x)\left[ (1-D)+D\cos (\frac{k\pi }{n})\right] \right|
<1,\;k=0,1,...,n-1.
\end{equation}
The more realistic case is to assume that the map depends on the agents e.g.
\begin{equation}
x_{i}^{t+1}=(1-D)f_{i}(x_{i}^{t})+\frac{D}{2}\left[
f_{i-1}(x_{i-1}^{t})+f_{i+1}(x_{i+1}^{t})\right] ,\;i=1,2,...,n.
\end{equation}
But analytic studies for Eq. (25) are more difficult.

These systems shed some light on how to control (synchronize) some CAS
[Ahmed et al 2003]. One may increase the coupling constant $D$. Also if the
network of the agents is more connected (e.g. SWN), then the system is
easier to synchronize. Finally external control can be applied preferably at
highly connected sites.

\section*{Appendix (5): Elements of multi-objective optimization}

Almost every real life problem is multi-objective (MOB) [Collette and Siarry
2003]. Methods for MOB optimization are mostly intuitive.\newline
\newline
\textbf{Definition (5):} A MOB problem is:
\begin{equation}
\mathrm{Minimize\;(min)}Z_{i}(\underline{x}),\;i=1,2,...,k,\;\mathrm{%
subject\;to}\;\underline{g}(\underline{x})\leq 0,\;\underline{h}(\underline{x%
})\leq 0.
\end{equation}
\newline
\newline
\textbf{Definition (6):} A vector $\underline{x}^{*}$ dominates $\underline{%
\acute{x}}$ if $Z_{i}(\underline{x})\leq Z_{i}(\underline{\acute{x}})\forall
i=1,2,...,k$ with strict inequality for at least one $i$, given that all
constraints are satisfied for both vectors.\newline

A non-dominated solution $\underline{x}^{*}$ is called Pareto optimal and
the corresponding vector $Z_{i}(\underline{x}^{*}),\;i=1,2,...,k$ is called
efficient. The set of such solutions is called a Pareto set.

Now we discuss some methods for solving MOB problems:

The first method is the lexicographic method. In this method objectives are
ordered according to their importance. Then a single objective problem is
solved while completing the problem gradually with constraints i.e.

\begin{equation}
\begin{array}{l}
\min \;Z_{1}\;\mathrm{subject\;to} \\
\underline{g}(\underline{x})\leq 0,\;\underline{h}(\underline{x})=0
\end{array}
,
\end{equation}
then if $\mathrm{ZMIN}(1)$ is the solution, the second step is
$\min \;Z_{2}$ subject to $Z_{1}=\mathrm{ZMIN}(1)$, and the
constraints in Eq. (26), and so on.

A famous application is in university admittance where students
with highest grades are allowed in any college they choose. The
second best group is allowed only the remaining places and so on.
This method is useful but in some cases it is not applicable.\\
\\
\textbf{Proposition (6):} An optimal solution for the
lexicographic problem
is Pareto optimal.\\
\\
\textbf{Proof.} Let $\underline{x}^{*}$ be the solution to the Lexicographic problem $%
P_{l}$. Thus
\begin{equation}
\underline{x}\neq \underline{x}^{*},\;\mathrm{then\;}Z_{i}(\underline{x}%
)=Z_{i}(\underline{x}^{*}),\;i=1,2,...,l-1\;\mathrm{and\;}Z_{l}(\underline{x}%
^{*})<Z_{l}(\underline{x}).
\end{equation}
Thus $\underline{x}^{*}$ is not dominated.\\

The second method is the method of weights. Assume that it is required to
minimize the objectives $Z(j),j=1,2,...,n$. (The problem of maximization is
obtained via replacing $Z(j)$ by $-Z(j)$. Define

\begin{equation}
Z=\sum_{i=1}^{k}Z_{i}w(i),\;0\leq w(i)\leq 1,\;\sum_{i=1}^{k}w(i)=1.
\end{equation}
Then the problem becomes to minimize $Z$ subject to the constraints. This
method is easy to implement but it has several weaknesses. The first is that
it is not applicable if the feasible set is not convex. The second
difficulty of this method is that it is difficult to apply for large number
of objectives. However it is quite effective for multiobjective problems
with discrete parameters since in this case Pareto optimal set is discrete
not a continuous curve.

The third method is the compromise method (sometimes called
$\varepsilon -$constr-aint method $P_{\varepsilon }(k)$. In this
case one minimizes only one
objective while setting the other objectives as constraints e.g. minimize $%
Z(k)$ subject to $Z(j)\leq a(j),\;j=2,3,...,k-1,k+1,...,n$, where
$a(j)$ are parameters to be gradually decreased till no solution
is found. The problem with this method is the choice of the
thresholds $a(j)$. If the solution is unique, then this method is
guaranteed to give a Pareto optimal solution.\\
\\
\textbf{Proposition (7):} If the solution is unique, then the
$\varepsilon -$constraint method is guaranteed to give a Pareto
optimal solution.\\
\\
\textbf{Proof.} Let $\underline{x}^{*}$ be the optimal solution
for the $\varepsilon - $constraint method then
\[
\forall \;\underline{x}\neq \underline{x}^{*},\;\mathrm{then\;}Z_{k}(%
\underline{x}^{*})<Z_{k}(\underline{x}),
\]
hence $\underline{x}^{*}$ is Pareto optimal. If $\underline{x}^{*}$ is not
unique, then it is weakly Pareto i.e. there is no $\underline{x}\neq
\underline{x}^{*}$ such that $Z_{i}(\underline{x}^{*})<Z_{i}(\underline{x}%
)\forall \;i=1,2,...,n$.\\

A fourth method using fuzzy logic is to study each objective individually
and find its maximum and minimum say $\mathrm{ZMAX}(j)$, $\mathrm{ZMIN}(j)$,
respectively. Then determine a membership $m(j)=(\mathrm{ZMAX}(j)-Z(j))/(%
\mathrm{ZMAX}(j)-\mathrm{ZMIN}(j))$. Thus $0\leq m(j)\leq 1$. Then apply $%
\max \{\min \{m(j),j=1,2,,n\}\}$. Again this method is guaranteed to give a
Pareto optimal solution provided that the solution is unique otherwise it is
weakly Pareto. This method is a bit difficult to apply for large number of
objectives. A fifth method is Keeney-Raiffa method which uses the product of
objective functions to build an equivalent single objective one.

\section*{Appendix (6): Fractional calculus in CAS}

Recently [Stanislavsky 2000] it became apparent that fractional equations
solve some of the above mentioned problems for the PDE approach. To see this
consider the following evolution equation

\begin{equation}
\frac{\mathrm{d}f(t)}{\mathrm{d}t}=-\lambda ^{2}\int_{0}^{t}k(t-\acute{t})f(%
\acute{t})\mathrm{d}\acute{t}.
\end{equation}
If the system has no memory then $k(t-\acute{t})=\delta (t-\acute{t})$ and
one gets $f(t)=f_{0}\exp (-\lambda ^{2}t)$. If the system has an ideal
memory, then

\[
k(t-\acute{t})=\left\{
\begin{array}{l}
1,\;\;\;t\geq \acute{t} \\
0,\;\;\;t<\acute{t}
\end{array}
,\right.
\]
hence $f\approx f_{0}\cos (\lambda t)$. Using Laplace transform

\[
L[f]=\int_{0}^{\infty }f(t)\exp (-st)\mathrm{d}t,
\]
one gets $L[f]=1$ if there is no memory and $L[f]=1/s$ if there is ideal
memory hence the case of non-ideal memory is expected to be given by $%
L[f]=1/s^{\alpha },\;0<\alpha <1$. In this case Eq. (28) becomes

\begin{equation}
\frac{\mathrm{d}f(t)}{\mathrm{d}t}=\int_{0}^{t}\frac{(t-\acute{t})^{\alpha
-1}f(\acute{t})\mathrm{d}\acute{t}}{\Gamma (\alpha )},
\end{equation}
where $\Gamma (\alpha )$ is the Gamma function. This system has the
following solution

\[
f(t)=f_{0}E_{\alpha +1}(-\lambda ^{2}t^{\alpha +1}),
\]
where $E_{\alpha }(z)$ is the Mittag Leffler function given by

\[
E_{\alpha }(z)=\sum_{k=0}^{\infty }\frac{z^{k}}{\Gamma (\alpha k+1)}.
\]
It is direct to see that $E_{1}(z)=\exp (z),\;E_{2}(z)=\cos (z)$

Following a similar procedure to study a random process with memory, one
obtains the following fractional evolution equation

\begin{equation}
\frac{\partial ^{\alpha +1}P(x,t)}{\partial t^{\alpha +1}}=\sum_{n}\frac{%
(-1)^{n}}{n!}\frac{\partial ^{n}[K_{n}(x)P(x,t)]}{\partial x^{n}},\;0<\alpha
<1,
\end{equation}
where $P(x,t)$ is a measure of the probability to find a particle at time $t$
at position $x$.

We expect that Eq. (30) will be relevant to many complex adaptive systems
and to systems where fractal structures are relevant since it is argued that
there is a relevance between fractals and fractional differentiation [Rocco
and West 1999].

For the case of fractional diffusion equation the results are

\begin{eqnarray}
\frac{\partial ^{\alpha +1}P(x,t)}{\partial t^{\alpha +1}} &=&D\frac{%
\partial ^{2}P(x,t)}{\partial x^{2}},\;P(x,0)=\delta (x),\frac{\partial
P(x,0)}{\partial t}=0\;\Rightarrow   \nonumber \\
P &=&\frac{1}{2\sqrt{D}t^{\beta }}M\left( \frac{\left| x\right| }{\sqrt{D}%
t^{\beta }};\beta \right) ,\;\beta =\frac{\alpha +1}{2}, \\
M(z;\beta ) &=&\sum_{n=0}^{\infty }\frac{(-1)^{n}z^{n}}{n!\;\Gamma
(-\beta n+1-\beta )}.  \nonumber
\end{eqnarray}

For the case of no memory $\alpha =0\Rightarrow M(z,1/2)=\exp (-z^{2}/4)$.

\end{document}